# A Survey on Intelligent Computation Offloading and Pricing Strategy in UAV-Enabled MEC Network: Challenges and Research Directions

Asrar Ahmed Baktayan[1], Ibrahim Ahmed Al-Baltah[1]

1 Department of Information Technology, FCIT, Sana'a University, Sana'a, Yemen

asrar@yemenmobile.com.ye, albalta2020@gmail.com

**Abstract** The lack of resource constraints for edge servers makes it difficult to simultaneously perform a large number of Mobile Devices' (MDs) requests. The Mobile Network Operator (MNO) must then select how to delegate MD queries to its Mobile Edge Computing (MEC) server in order to maximize the overall benefit of admitted requests with varying latency needs. Unmanned Aerial Vehicles (UAVs) and Artificial Intelligent (AI) can increase MNO performance because of their flexibility in deployment, high mobility of UAV, and efficiency of AI algorithms. There is a trade-off between the cost incurred by the MD and the profit received by the MNO. Intelligent computing offloading to UAV-enabled MEC, on the other hand, is a promising way to bridge the gap between MDs' limited processing resources, as well as the intelligent algorithms that are utilized for computation offloading in the UAV-MEC network and the high computing demands of upcoming applications. This study looks at some of the research on the benefits of computation offloading process in the UAV-MEC network, as well as the intelligent models that are utilized for computation offloading in the UAV-MEC network. In addition, this article examines several intelligent pricing techniques in different structures in the UAV-MEC network. Finally, this work highlights some important open research issues and future research directions of Artificial Intelligent (AI) in computation offloading and applying intelligent pricing strategies in the UAV-MEC network.

## 1   Introduction

Mobile Edge Computing (MEC) has emerged as a promising model by incorporating mobile network edge telecommunications and cloud computing services to provide wireless access network computing and communications capabilities near mobile devices (MD) [1]. Besides, computation offloading was initially implemented as a promising paradigm for improving the computational capabilities of devices by enabling them to use remote cloud servers to perform their computing tasks. Computational offloading to remote cloud servers could therefore speed up computational tasks, but you would not be able to meet the latency requirements due to high association delays. Computational offloading paradigms suggest moving computing and storage resources closer to end-users, i.e., to the edge of the network, to meet the extremely low latency requirements of delay-sensitive emerging applications such as Augmented Reality (AR), Virtual Reality (VR), and autonomous driving [2]. It is possible to loosely split the research literature by allocating resources into two groups. In order to fully use the available resources, the first category is to follow a rational allocation of computing resources [3] [4] . The second group is the implementation of auxiliary nodes to increase the computational capacity of the MEC system [5]. Besides, sharing the computational load of the MEC, implementing the auxiliary nodes may also allow the use of idle resources [6]. These auxiliary nodes may be Device to Device (D2D) communication [7], remote clouds like Unmanned Aerial Vehicles (UAVs) with computational capability [1] [8], and MEC servers in adjacent areas.

MEC networks operated by UAVs have the ability to improve computing performance and reduce execution latency. In addition, UAVs are used as a relay edge computing node, and UAV-enabled MEC networks are proposed as a solution to the current MEC network's shortcomings, which include fixed base stations (BSs) and limited computing power nodes [9]. Instead of deploying additional fixed communication networks, it is highly productive and cost-effective to temporarily deploy a UAV as an aerial BS to provide cellular services in the heavy traffic regime. In contrast to the more static terrestrial alternative, the essence of UAV-MEC activity is more harmonious with the complex and transient changes in the environment [10]. Because of the high speed and varying mobility patterns in a multi-UAV-MEC network, UAV-MEC environments need highly accurate data localization with shorter time intervals [11]. The infrastructure-based topology of UAV-MEC aided MNOs depends on the current network BSs. As a result, UAV-assisted MNOs follow the aided networks' wireless protocols (e.g., Mobile Cloud Networks (MCNs) and WLANs). UAV-MECs may use UAV-aided communications networks to expand the communication network to areas with weak connections. In this regard, the Air-to-Air (A2A) link establishes communications between multiple UAV-MECs. A2A's architecture aims to ensure that collaborative communication efficiency in a multi-UAV-MEC network is maintained. Furthermore, A2A can be easily modelled as a line-of-sight (LoS) propagation, and the connection quality of A2A is vulnerable when multiple UAV-MECs are moving. Hence, the focus of the investigation shifts from improving the A2A connection quality to collaborative role scheduling of multiple UAV-MECs. Since UAVs move about, the locations, in which they collaborate are constantly changing, necessitating the use of routing protocols [12].

In reality, the UAV-MEC is not able to control the sum offloaded by MDs and the MDs' offloading benefit capacities are imperceptible to the UAV-MEC. Other than that, MDs are unwilling to report their data to UAV-MEC servers like resource capacities and battery states since these data typify as a private data. As a result, offloading is a critical thought in optimizing resources, since MDs and tasks are not homogeneous, and diverse MDs have diverse costs or negative valuations towards holding up time [13]. Given that the UAV-MEC servers' resources are shared by selfish MDs, pricing may be a valuable tool for spurring MDs to internalize the negative externality of the delay they cause to other MDs. Hence, the MNOs and content providers discover themselves connected in this showcase profit generating, and subsequently in competition for caching resources in the shared space possessed by multiple MNOs and setting cost issues [14]. As networks became increasingly crowded, the use of new techniques to design future systems had to be basic. Hence, the development of Artificial Intelligence (AI) and next-generation systems are placing with their focus on pricing matters. It is not enough to adapt these elective models to the current organization to make progress on future systems; we do not have to apply AI and its creativity only, but we also need to improve the organized network architecture [15].

Despite the existence of plenty of related offloading and pricing studies for UAV-enabled MEC in the literature, the survey [16] reviewed many applications of economic and pricing models in Wireless Sensor Network (WSN) and machine-to-machine (M2M) communication. In addition, work [17] produced a systematic, informative, and thorough survey paper focusing on stochastic-based offloading processes, including profit-making of offloading for MDs and providers. Similarly, conventional pricing models may not be effective or applicable for 5G networks, so economic and pricing models in [18] are proposed to address the resource management problems in 5G wireless networks, including MD association, allocation of bandwidth, interference, and power management. Furthermore, the survey [19] focused on the role of prices in communication networks, depending on the degree of strategic interaction between MDs themselves and between MDs and MNOs, and analyzed their effect on the performance of the network in various scenarios. An antenna arrays, signal management, and the use of centralized and decentralized techniques for UAV communication systems in survey [9] looked at UAV communication systems for both hardware and algorithm-based applications.

The survey [20] looked at previous studies on computation offloading models. The authors looked at addressing a variety of offloading simulation models based on several paradigms, including Markov decision process (MDP) and Reinforcement Learning (RL), as well as game theory. In terms of design aim, performance, special features, features, and offloading methodology, the authors of [21] investigated the offloading algorithms in UAV-MEC. The work [22] is for wireless network pricing, which addresses the activities of two characters from an economic point of view in the D2D data exchange model: data buyers who want to purchase a certain amount of data and data sellers who want to sell data through the D2D network. The authors of [23] described and categorized the current game-theoretic methods used in wireless communication with UAVs. The main technological problems are highlighted, followed by a summary of the game-theoretic strategies used to address them. They presented a case study to illustrate the advantages of applying game theory. They also addressed the issue of interference and suggested a mean field game (MFG). In order to address issues with energy, security, task offloading, and latency of IoT data in 6G networks, the survey [24] concentrated on the usefulness of UAV computing and the crucial role of Federated Learning (FL).

Table. 1 summarizes recent surveys on computation offloading in UAV-MEC and AI that are especially relevant, and compares their coverage to key areas of this survey paper.

Table 1. Existing surveys on computation offloading in UAV-MEC

| UAV survey | UAV enabled MEC | AI solution | Computation offloading | Offloading challenges in UAV-MEC | Price for UAV-MEC offloading | Description |
|---|---|---|---|---|---|---|
| [23] | √ | √ | × | × | × | understanding of UAV-aided networks architecture, benefits, challenges, and various game theoretical solutions |
| [24] | √ | √ | √ | √ | × | intelligent UAV computing offloading to enable 6G networks |
| [25] | √ | √ | × | × | × | surveys game-theoretic and machine learning algorithms in UAV networks |
| [26] | √ | √ | √ | √ | × | describe UAV-RAN architecture and fundamental features for the development of 6G networks |
| [27] | √ | × | √ | √ | × | 3GPP standardization and emphasizes socio-economic concerns in UAV networks |
| [28] | √ | √ | √ | √ | × | UAV-MEC intelligent computing with various optimization objectives. |

| [29] | √ | × | √ | √ | × | basic terminologies, solutions and architectures used in UAV-MEC networks for offloading |
| [21] | √ | √ | √ | √ | × | UAV-enabled MEC solutions in computation offloading |
| [30] | √ | √ | × | × | × | edge AI on the UAV technical aspects |
| This paper | √ | √ | √ | √ | √ | Intelligent computation offloading and price strategies in UAV-MEC |

However, further study and debate are required before using UAV-MEC to assist cellular networks for computation offloading using AI, and dynamic pricing strategies (using AI methods) for resource allocation. This paper offers a comprehensive overview of the current state of UAV-MEC-assisted cellular networks. To the best of our knowledge, there is a serious lake for a comprehensive review that should focus on the intelligent computation of offloading and intelligent pricing strategies in UAV-MEC networks using AI to offload data from MDs to UAV-MEC operators. Therefore, this paper provides some contributions to advance the current state of the art. Firstly, it attempts to create a panoramic view about the intelligent computation offloading, and intelligent pricing strategies in UAV-enabled MEC networks through studying a noticeable number of related published researches. Secondly, it provides a comprehensive classification and review for the intelligent techniques that have been proposed for data computing offloading in UAV-MEC. Thirdly, present the pricing strategies in UAV-MEC. Finally, it points out some challenges issues and research directions for further improvements to advance the current body of knowledge.

The rest of this paper is organized as follows. Section 2, the explanation of offloading process in UAV-MEC and the factors affecting data offloading including pricing. In Section 3, the intelligent UAV-MEC models for computation offloading are analysed. Then, the intelligent pricing decision of computing offloading in UAV-MEC network. Specifically, in Section 4, the intelligent pricing strategies in UAV-MEC are discussed. In Section 5, the challenges and open issues are discussed. Finally, conclusions are given in Section 6.

## 2   UAV Mounted MEC in 5G for Intelligent Data Offloading

The rapid advancement of technology has resulted in the use of a wide range of UAVs in different applications in recent years. UAVs have been widely used in social applications such as communication, detection, and rescue due to their tiny size and high mobility [31]. A drone, or UAV, is an aircraft that does not have a human pilot on board. UAVs are one part of an Unmanned Aircraft System (UAS), which includes the UAV, a ground-based controller, and a communication system between the two. UAVs were originally created and deployed for military purposes, but there has been an increase in drone usage in 5G networks research in recent years [32]. As a result of its high flexibility and mobility, the UAV is widely used as an aerial MEC node for service coverage in surveillance, data collection, disaster assistance, and public safety. MEC enabled by UAVs connects MDs to remote resources, but due to the restricted embedded batteries, it still confronts design constraints for communication and computing [33]. When interacting and working with other cellular technologies, UAVs have good sensing skills, resulting in intelligent outcomes that give more value to MDs [34]. The ambitious requirements of 5G and beyond 5G (B5G) wireless networks have recently envisioned a larger range of targets in terms of greater coverage and connection, multiple slices are designed to be intelligently placed in the same physical infrastructure to accommodate different types of services with different quality of service (QoS) needs [35] [36].

As moving MDs distribution changes dynamically in real time, the deployment of static/fixed edge computing nodes in cellular networks fails to meet their computing demands [37]. As a result, cellular communication with the assistance of UAV could be great enhanced and more efficient, which is being investigated by academia and industry due to the less conveyed framework and arrangement in a 3D space that can offer assistance systems with ever-increasing user necessities and scenarios. Given capabilities such as quick mobility and adaptability, aerial base stations (ABSs) are an imperative building piece for the Next Generation Network (NGN) [38]. The future roles of ABS frameworks in 5G systems and B5G are to provide network on request as standalone airborne networks or to help the current terrestrial cellular networks. Drones (UAVs) can help ground systems for numerous purposes, such as offloading MDs data from ground stations to upgrade the capacity of existing ground networks [39]. Drone assistant networks provide a fast, effective, and cost-effective arrangement. Each MD is served by either the interconnected UAV-MEC server within the same cell or the UAV-MEC server found in another UAV-MEC, as long as the delay from MD to serving UAV-MEC server meets the service delay necessities. Therefore, the UAV-MEC server that is serving the MD cannot be too far from the MD, i.e., the delay spread between them should be limited. However, the quality of communication is hard to guarantee due to the obstruction of dense buildings, lack of infrastructure in some zones, or even the LOS of mmWave. Therefore, UAVs have established communication links between the two ends due to their characteristics of ignoring the terrain and flexible deployment [40].

## 2.2 Offloading Process in UAV-MEC

A UAV-MEC-based offloading process consists of three relatively independent processes, i.e., profiling, partitioning, and decision-making [41]:

1. **Profiling**: The framework should have up-to-date information on the status of the application and the MD in order to make accurate offloading decisions. Application profiling is the practice of gathering information on programs, such as energy consumption, data size, memory usage, and execution time. MD profiling, on the other hand, collects data on MD status and may not be sent to edge computing, such as battery level, CPU utilization, wireless connection [41], and sojourn time [42]. The sojourn time is the duration of the stay of an MD in UAV-MEC coverage. It denotes the amount of time available for offloading before leaving this coverage area, whereas the transition time between two UAV-MECs indicates how long MDs will wait for feedback [43]. The authors of [43] proposed a BS selection method based on the MD's time spent in the selected network. They also demonstrated a relay mechanism for reducing the danger of task loss owing to MD motion during the offloading phase. The relaying technique treats task volume and application type as random variables, allowing for greater flexibility in the solution.
2. **Partitioning**: Based on diverse facts, application partitioning divides the task into offloadable and non-offloadable components. Because the non-offloadable segment may require equipment that is not available at the offloading location, such as a camera, GPS, or accelerometer, it must be performed locally on the MD [44].
3. **Decision-making**: The following four factors must be considered when making an offloading choice. To begin, the selection of which parts of fine-grained tasks are to be offloaded and which ones are to be executed locally. Second, the determination of the best node to offload. Third, the best moment to offload must be determined, such as when the network environment is favourable, when the amount of transmission data is minimal, or when the amount of processing is substantial. Finally, a transmission method must be chosen, such as WiFi or a cellular network [45].

## 2.3 Factors Affecting Intelligent Offloading Decisions

In dynamic environments such as 5G networks with UAVs, knowing whether to undertake this computational offloading is a mission of high importance and complexity that requires intelligent decision-making [46]. Decision on the task offloading mechanism is a choice on whether to offload a task or not, taking into account parameters such as MD battery power, memory space, UAV speed and trajectory, and wireless transmission latency, as well as ensuring that the required QoS is verified when MD roams to a closer cell [47]. Table 2 illustrates some factors that influence the offloading decision. In the following, some details about the factors are discussed.

Table 2. Factors affecting the offloading decision

| Factor | Value |
| --- | --- |
| Application type | delay-sensitive, delay-tolerant |
| Wireless environments | 4G;5G;WiFi |
| User preference | Data confidentiality, privacy |
| Mobile device specification | CPU speed, available memory, battery level |
| Server specifications | CPU speed, available memory, load |
| Mobility | Distance between MD and server |
| Price | Cost of computation and communication and operator profit |

**Application type**: Varying types of applications place different weights on the two criteria of delay and energy usage. To attain optimal performance, the offloading decision must take into account the various variables. There are two types of applications. Maximum delays of minutes to hours are tolerable for some mobile applications that handle machine learning model training and personal health analytics. When edge resources are scarce or expensive, offloading these workloads to powerful public cloud servers for processing is a possible option [48]. New participative sensing applications are a great example of big data delay-tolerant applications. As a result, for delay-tolerant applications, response time is not the most important consideration; instead, lowering energy usage is. The authors of [49] proposed a handover (HO) choice method that takes into account both radio and computing offloading. The proposed HO decision algorithm works with femtocells to reduce the effect of radio congestion by offloading static data. It also collaborates with the remote cloud to offload delay-tolerant processing, decreasing MEC congestion.

For some applications that require a certain amount of time (e.g., natural language processing, face recognition, virtual reality, vehicular communications), subscribers to mobile services expect a response time that is comparable to their cognitive capacity. In some situations, such as the usage of autonomous automobiles, slight timing errors can result in serious security or fatal issues. As a result, for a better user experience, these applications' reaction times should be as short as possible. In [50], deep learning was used to figure out the best task partition and offloading policy for a certain task. The deep neural network (DNN) is trained using a big dataset obtained from a mathematical model, which reduces the time delay and energy consumption of the entire process, satisfying the needs of delay sensitive applications.

**Wireless Environments**: For data transfer, mobile devices frequently use several wireless interfaces with different availability, delay, and energy costs, such as 3G, 4G, 5G, and WiFi. The data rates that different access technologies may achieve is varying greatly depending on the environment. Long-term evolution advanced (LTE-A) can deliver 3Gbps in DL and 1.5Gbps in UL for 4G. WiFi speeds of up to 600 Mbps are possible with the 802.11n wireless local area network standard. The introduction of the 802.11ax standard increased the maximum data bandwidth to about 9.6Gbps [51].

**Computation and Communication Resource Allocation**: Some offloading-related actions must be considered when an offloading decision is made. For example, for the transmission and processing of each task, an appropriate amount of communication and computing resources must be allocated.

The 5G networks deal with a wide range of devices that must use the required resources in a selfish manner [52]. Together with MEC and 5G, task execution delay and download time delay would be reduced, and MD QoE might be increased. In real-world scenarios, a MEC server's computation resource is limited to some extent. As a result, preventing MEC resource abuse and properly allocating MEC resources becomes a vital issue for a MEC system [53]. In contrast to cloud computing, edge computing places computing and caching resources closer to edge devices, reducing latency and increasing data rate while also lowering energy consumption [54]. Due to network infrastructure is being limited, it is critical to maximizing network utilization by allocating resources to different services based on their requirements to maximize resource allocation and meet the required SLA [55].

The calculating performance of computer equipment is controlled by the CPU cycle frequency in general (also known as the CPU clock speed). In the state-of-the-art mobile CPU design, the sophisticated dynamic frequency and voltage scaling (DVFS) technique was utilized, which allows adjusting the CPU-cycle frequency (or voltage), resulting in increased or decreased energy consumption. The complexity factors, or input data size and computational demand (CPU cycles per bit), are the most critical aspects of task processing. These elements are essential for understanding the varied computing requirements. Applying filters to an image, for example, uses fewer CPU cycles than applying a face detection method to a video [56]. The MDs require resources from the edge cloud to fulfil computation duties, but the edge cloud must keep its available CPU cycles for processing the entire offloaded data below its calculation capacity. As a result, the edge cloud is thought to price CPU cycles to control processing resource demand and supply. The MDs then divide their input data for local computing and offloading according to the prices set by the UAV-MEC server [57].

To leverage the DVFS capability, the back-pressure algorithm was used in [58] to determine the offloading decision and the number of tasks that might be offloaded, and the authors presented a computational offloading technique for reducing latency. The authors built a system with several local MDs work queues and numerous edge processor task queues. The authors then developed an offloading approach based on minimizing the Lyapunov drift optimization problem to reduce work queue length in each time slot. The amount of energy required to convey a given amount of data might vary by an order of magnitude or more. The WiFi interface is, on average, more energy efficient than the cellular interface. Because it uses ultra-high-frequency bands (e.g., millimetre-wave), which have smaller beams and weaker diffraction ability, 5G consumes the most energy [59].

Not only can BS availability and quality vary by location, but uplink and downlink bandwidths also fluctuate according to a variety of circumstances such as weather, mobility, and building shields. When there is good connectivity, data transmission uses a lot less energy than when there is not [60].

**User Preferences**: Combining a UAV with MEC can give real-time services. Attackers, on the other hand, can easily get access through the use of UAVs, MDs, and intelligent equipment. As a result, processes are hampered, and data security is compromised [61]. The study [62] presented a secure UAV-MEC system, in which many ground MDs offload enormous computational chores to a close real UAV in the situation of many snooping UAVs with incorrect placements. To strengthen security, both full-duplex genuine UAVs and non-offloading ground MDs provide jamming signals.

**Mobility**: The coverage changes since the UAV-MEC and MD are moving. The UAV-MEC should find an optimal hover point and determine the coverage at a fixed height to service additional MDs while limiting its energy consumption [63]. Authors of [64] created a virtual cell, in which each active vehicle user is served by a large number of low-power BSs. Each moving MD is considered an agent. A non-cooperative stochastic game for radio and electricity resource allocation was developed and solved by leveraging multi-agent RL efficiency, taking into account practical location-aware node relationships.

**Price of UAV-MEC**: The valuation is the highest price a MD is willing to pay for commodities, as well as the MDs' preference for resources. This is a parameter that only the MD has access to. The price algorithm is meant to prevent MDs from cheating. The MEC server calculates each MD's actual payment and returns the results of offloaded duties to the associated MD [65].

By deploying parked vehicles as temporary computation service providers, Yangzhe Liao et al. [66] presented a multi-user vehicle-aided multi-access MEC network architecture. A dynamic pricing technique is provided to minimize MD energy usage while optimizing mobile MNO revenue under quality of service restrictions. To identify MD's fine-grained offloading decision, a differential evolution approach is proposed. In addition, a Q-learning algorithm is used to determining the appropriate unit service price.

In the same context, authors of [67] formulated the Small Service Providers (SSPs) the problem of determining their traffic-dependent prices as a Bayesian game, and they first showed that there exist no Nash equilibriums in pure strategies. They then proceed to derive the structure of a mixed-strategy symmetric Bayesian Nash equilibrium (BNE). They also compared the flexible pricing scheme with the traditional flat-pricing scheme (where the SSPs are restricted to announce a single price, irrespective of the traffic that is offloaded to them) in terms of the payoffs achieved by the SSPs as well as the MNO.

The operators of the UAVs in [68] must determine the best beaconing period (the time it takes to send brief signals announcing the presence of a UAV) while conserving energy. They also determine the optimal pricing approach to optimize their respective market share by encouraging MDs to offload their data to the UAV. As a result, both in terms of pricing and availability, a tractable analysis for the game's Nash Equilibrium is derived. See table 3 for more work about factors affecting offloading. Table 3 shows offloading modes and purposes.

Table 3. Offloading Modes and Purposes

| Ref | Offloading Purpose | UAV structure | Device type | Offloading mode |
|---|---|---|---|---|
| [69] | Time sensitive data | One UAV | IoT cluster | Partial (sequential) |
| [33] | Time sensitive data | One UAV | IoT | Partial (sequential) |
| [70] | Time tolerant | Multi-UAV | vehicles | Partial (sequential) |
| [71] | WiFi with cellular | One UAV | users | Partial (sequential) |
| [72] | Privacy | One UAV | IoT | binary |
| [73] | Time sensitive data+ UAV battery | Multi-UAV | IoT | Partial (sequential) |
| [74] | Time sensitive data + User battery | Multi-UAV | IoT | Partial (parallel) |
| [75] | Mobility + user and UAV battery | One UAV | Mobile user | Partial (sequential) |
| [76] | Price + vehicle and UAV battery+ time sensitive | One UAV | vehicles | Partial (sequential) |
| [77] | Time sensitive data | Multi-UAV | MDs | Partial (parallel) |
| [78] | Data confidential + Price | One UAV | IoT | Partial (sequential) |
| [79] | Task completion of MDs | One UAV+ one ground BS | MDs | Relay |

As shown in table 3, the offloading paradigm is given in UAV-MEC when MD prefers to run the task to the UAV-MEC server. Partially offloading means that the MD's tasks are partially offloaded to the UAV-MEC server and partially executed locally for a better customer experience [80]. There is also relay offloading when the UAV acts as a relay to connect MD to ground BS or ground MEC [81]. In general, partial task computing in UAV-MEC can be done in two ways: sequential one by one or parallel offloading multiple tasks at the same time.

## 3 Computation Offloading Utility in UAV-MEC

The network Utility is wisely transmitting computations over various devices that share network resources [82]. It is not possible to collect all optimization factors in one node and calculate the optimization problems in a centralized manner because the number of MDs and connections to the UAV-MEC will grow exceptionally high in the 5G network. Thus, the computation offloading problem needs to be solved in a distributed manner by online algorithm calculations. Decomposition techniques are one of the relevant mechanisms for understanding the distributed computation problem. The computation offloading problems have been presented in the past as deterministic environments. Notable progress has been made in improving the utility concept by taking into account the stochastic existence of the network and MDs. The common concept of the network utility problem includes 1) an objective function comprising all utility functions, each connected to a specific MD (or flow) and 2) A collection of imperatives that fundamentally define a communication network's limitations. The goal then is to prepare a network protocol that is close to a distributed solution approach and achieves the ideal allocation of rates. The concept of utility is closely linked to the concept of Quality of Service (QoS) and Quality of Experience (QoE); the previous speeches are about a technique for measuring the QoS provided to MDs, while the degree of MD satisfaction with a specific service is calculated in the last listed speeches. The three definitions are, therefore, closely related [83].

The MEC network, which includes UAVs and central clouds, is appropriate for a variety of complex application scenarios. After approaching the MD along the predetermined path, the UAV hovers. At that point, UAV now acts as a fog node (FN) in the UAV-MEC network, serving MD [84]. The UAV-MEC can undertake two functions of task computing offloading or relay forwarding. Task computing offloading and relay forwarding are two functions that the UAV can perform [85]. As the network's middle layer, UAV-MEC collaborates with the central cloud to process tasks with offloading and caching [86]. The network utility metrics consider both objectivity and total data rate in a multi-UAV-MEC network more than the pure data rate metric. Co-channel interference can be efficiently improved by correctly designing the positions of UAVs-MEC and MD associations, thereby increasing the network utility. It is worth remembering that the network utility model was used to assess fairness in several resource-allocation problems [87].

The 5G networks deal with different devices that need to selfishly use the resources required. Thus, because of the limited network infrastructure, it is necessary to ensure the effective use of networks' resources by allocating resources to various services in order to optimize the allocation of resources and comply with the required SLA [55]. In particular, when fewer MDs are competing for the computation resource of the UAV-MEC server, an MD is more likely to offload its data to the UAV-MEC server with no congestion, or when a strong channel status (for higher rate and better QoE) or low UAV-MEC service price [88]. However, there are some obstacles that remain to hinder the full use of the benefits of computation offloading in UAV-MEC networks [89].

Wireless caching refers to caching content in the storage/memory of small base stations (SBSs) integrated with MEC as UAV; such content is like common video or web information. The offloading of MD data enables the use of UAV-MECs to minimize traffic on Macro-Cell Base stations (MBSs). Such services can (1) reduce the transmission latency of MD content requests [90], (2) save the battery of the MD [91], (3) Reduce the redundancy of common content transmissions over backhaul links that provide greater protection [92], (4) reduce transmission costs [93], (5) enhance performance [94] [95], (6) achieve greater energy efficiency for MDs, UAVs, and MNO, (7) greatly enhance the ability of the network to support time-sensitive applications [96] [97] as well as (8) improve coverage [98] [99], see fig.1. The UAV-MECs and MBSs, however, may belong to various owners who have distinct interests. In addition, UAV-MECs usually have limited resources, i.e., power, bandwidth, and storage [18].

The computational offloading of the decision-making problem of MDs is typically conceived as a socially aware computational offloading game based on game theory [100] or the multi-armed bandit (MAB) theory [101], where each MD aims to maximize its utility in the social group, which is the total of its individual utility and the weighted sum of the group utilities. Recently, AI and Machine Learning (ML) based approaches such as deep learning and DRL have an estimated optimal solution such that each MD can decide its decision to offload to UAV-MEC without knowing the offloading decision of other MDs [102] [103].

The efficient negotiation of pricing is one of the key factors of computing offloading. It can help increase the profitability of UAV operators and MNO, balance the usage of resources, and enhance the QoE of the MD. Determining the "right" price, however, is a very complex decision-making issue that dilemmas stockholders, as it needs to take into account information from many MDs demanding task offloading [104].

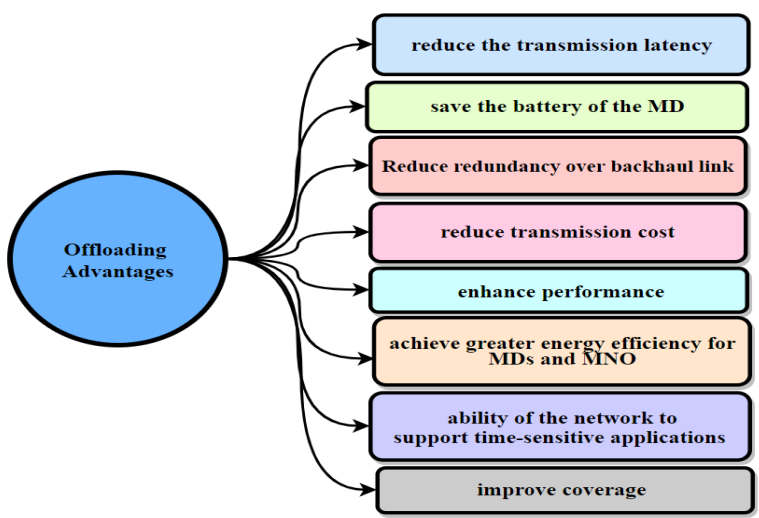

Figure 1. Advantages of offloading

## 3.1 Intelligent Models for Computation offloading in UAV-MEC

Integer programming, game theory, and the Markov Decision Process (MDP) are the three key types of mathematical models for multi-UAV-MEC cooperative task assignment currently available. Different models represent various approaches to the problem. In general, the integer programming model is the most intuitive to construct, and it often seeks to optimize system benefits. Game theory models, on the other hand, begin by optimizing the individual utility of each UAV-MEC, making them more efficient in situations where communication is minimal. The MDP, on the other hand, is more concerned with the ambiguity posed by the environment of UAV-MECs. There are two kinds of algorithms: centralized and decentralized algorithms, based on the communication structure between UAV-MECs themselves and with the global entity. Computational complexity is often a major concern when solving these problems in order to achieve better feasible solutions at a faster speed [105]. When the human pilots a swarm's activity by manipulating a leader drone, the other drones obey the leader autonomously based on the intensity of cellular signals. The use of UAV-MEC swarms for general applications has recently begun to gain popularity. There have been numerous UAV-MEC swarm demos, but the degree of autonomous operation in most of them has been limited [9] Fig.2 illustrates the intelligent algorithms for computation offloading in UAV- MEC.

### 3.1.1 Convex and Non-Convex Optimization

Convex optimization is a powerful strategy for tackling optimization problems because it is solvable. In a convex model, the offloading goal is expressed as an objective function, and the offloading constraints are expressed as constraint functions. Traditional approaches, such as the Lagrange duality method, can be utilized to solve the specified optimization model and meet the global optimization goal if it is convex. Converting a non-convex optimization problem to a convex optimization problem is a typical solution if the offloading model is non-convex. Non-convex problems are extremely difficult to solve because they are NP-hard [106]. There are several ways for transforming a nonconvex problem into a convex problem that may be used to discover a solution in polynomial time, such as the reformulation linearization technique and the Successive Convex Approximation (SCA) methodology [107]. In the paper [108] the authors broke down the problem of finding the best partial offloading scheme for multiple nonlinear programming problems (NLPs) and suggest an effective algorithm for solving them using block coordinate descent (BCD) and convex optimization techniques. On the other hand, to tackle the non-convex problem, the work [109] is investigated the UAV-enabled wireless controlled cooperative MEC system, in which a UAV equipped with an energy transmitter and a MEC server provides both energy and computing services to sensor devices (SDs). The active SDs chose to finish their computing tasks with the aid of the UAV and their nearby idle SDs who are not working on something. Moreover, deep learning Neural Networks (NN) are universal function approximators [110]. They can accurately approximate any function if there are enough neurons. This is the attraction of these kinds of systems. Non-convex functions are becoming more popular as approximation functions. The high execution time and energy consumption of deep learning services on lower-end devices continue to be substantial roadblocks to their widespread use. In [111], a deep learning network is trained to calculate the remaining energy and choose the best offloading technique.

### 3.1.2 Game Theory

The game theory model portrays a game between player groups that choose to behave cooperatively or non-cooperatively in order to maximize their benefits (payoffs) by implementing the strategies through the collective behaviour of the players. In the same context, the computation offloading game theory can be divided into cooperative and non-cooperative games as shown in Fig.2. Collaborating nodes with the aim of improving the security of the entire network against various security threats define cooperative games. Non-cooperative games, on the other hand, entail competing for individual acts, in which each node attempts to maximize its own payoff at the expense of the others' outcomes [23]. Game theory is a good tool for building distributed processes. It can be utilized in a multi-user offloading scenario, where each MD chooses a suitable method locally to arrive at an offloading solution that is mutually acceptable. Each MD makes a decision on offloading, gains incentives, and updates the decision. This process is repeated until the incentives can no longer be increased, i.e., until Nash equilibrium is reached.

#### a. Cooperative Game algorithm for Computation offloading

-**Bargaining game**: Two or more players with competing interests must agree on certain tactics that lead to a collection of solutions in a bargaining problem. In contrast to non-cooperative UAV-MECs, cooperative UAV-MECs aim to produce better performance [112]. The bargaining solution in [113] has two parameters namely α and β. The first parameter α can be interpreted as the reference point's importance, while the second parameter β can be interpreted as the reference point's ability to form players' goals. The (α /β) bargaining solution has a variety of interesting axiomatic properties from a game-theoretic standpoint. This approach can be adaptively implemented to provide a fair efficient solution for the task partitioning problem based on the infrastructure of Flying Ad-hoc Networks (FANETs). As a result, the authors made control decisions during the bargaining game process to achieve a globally acceptable system efficiency while optimizing UAV profits. Similarly, to find the best fit between UAVs and MDs, the authors of [114] developed a matching mechanism based on both UAV and MD preference lists. To improve offloading performance and optimize the utility of UAVs and vehicles, the transaction technique of calculating data between UAVs and MDs is then treated as a bargaining game.

-**Coalition Formation Game (CFG):** Coalition game theory is a branch of game theory that explores how players act when they cooperate and offers a basis for clustering. It has been commonly used in wireless communication and signal processing, and it's given researchers new insights into task allocation, interference management, and power control, among other things [115]. For correlative tasks, multiple UAVs form groups and communicate with one another. If UAV-MECs belong to the same operator, they must follow the operator's order. Win-win situations among drones require inter-coordination in UAV-MEC coalitions, even if they come from different operators. As the exchange process occurs, it increases the local services of the selected UAV-MECs, which are the same as the path with overall utility. Other coalitions decide the local usefulness of UAV-MECs that are not in the relevant coalition. It should be noted that changing the strategy of a single UAV has a direct impact on the entire system, so the system's utility will converge to the optimal state [116]. In coalition-based UAV-MEC swarms, the authors [117] investigated the joint deployment, computation offloading, power management, and channel access optimization issues. A random best and better response algorithm is proposed to solve the problem of satisfying the distributed nature of UAV-MEC swarms. In [118] the authors looked at FANETs from the standpoint of intelligent optimization, proposing a coalition-based networking architecture for UAV internal communication. UAVs are divided into several coalitions in the proposed model to handle various coordination subtasks including data collection and reconnaissance. In this regard, the ground nodes in [119] actively form multiple coalitions in response to the UAV's location, and the UAV changes its position in response to the data distribution of ground networks. A Stackelberg game is used to model the relationship between the UAV and the ground nodes. For data collection among ground nodes, a coalition formation game (CFG) is created. The proposed CFG an exact future game with at least one Nash equilibrium.

**b. Non-Cooperative Game for Computation offloading in UAV-MEC**

- **Stackelberg-Game**: The Stackelberg game's basic strategy is that one character, referred to as the chief, has the right to take the first action. The other player, known as the follower, then watches the leader's actions and makes its own decision based on what it observes. In a software-defined network (SDN)-based Vehicle Ad-hoc Networks (VANETs) [120], authors suggested a Stackelberg-game-based pricing scheme for data offloading, which finds the best price and offloading percentage to make data distribution from service providers to service requesters easier. Unlike other approaches, this scheme uses moving vehicles as service providers and takes into account their specific characteristics, such as vehicle mobility, Vehicle-to-vehicle (V2V) communication length, BS bandwidth, service provider, and service request popularity, and content popularity. If a moving vehicle requires content, the lowest-priced neighbouring vehicle is chosen as the service provider, allowing the vehicle to receive a portion of the content from the BS and the remainder from the service provider. To model the relationship between a service provider and a service requester, the Stackelberg game is used.

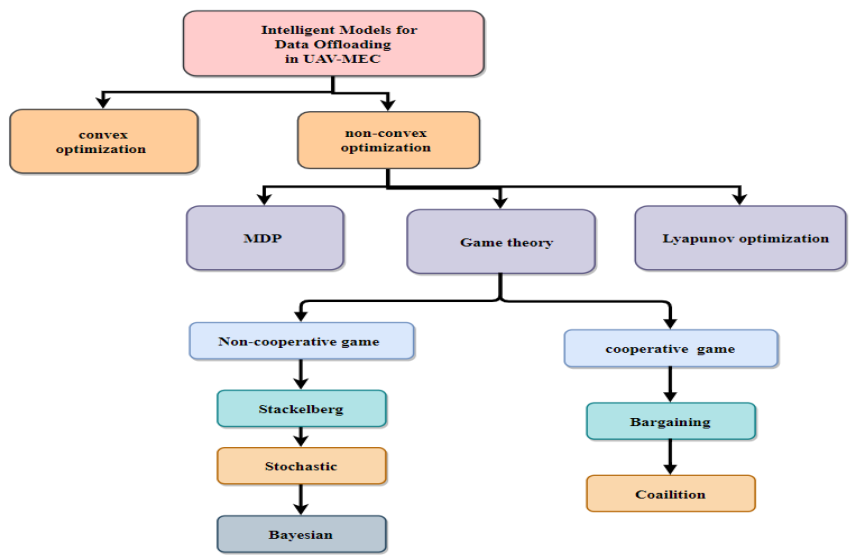

Figure 2. Intelligent algorithms for computation offloading in UAV-MEC

- **Stochastic game**: The advantages under the current state are split into two parts in the stochastic game model. One part is the immediate benefits after making a decision in the current state. Another aspect is the potential indirect advantage in the next state. Both the current state and the next state decide the gains of the current state. For illustration, the long-term resource allocation problem in [121] was conceived as a stochastic game with the goal of optimizing expected rewards, in which each UAV serves as a learning agent and each resource allocation solution corresponds to an action taken by the UAVs. Following that, they created the multi-agent RL system, in which each agent uses learning to discover the best strategy based on its local observations.

- **Bayesian game**: A Bayesian Nash equilibrium (BNE) is one of the potential game solution principles. In other words, no player in the BNE assumes that deviating from their pure strategy would increase their anticipated stage payoff. The authors of [122] demonstrated that the best response strategies of followers (terrestrial base station and UAV) in perfect Bayesian equilibrium (PBE) are determined by a solution of the partially observable Markov decision process (POMDP). They also created a Bayesian RL (BRL) algorithm that allows each follower to update its beliefs about the POMDP's unobservable states and predict its strategy based on experiences with other followers and with the environment.

### 3.1.3 Lyapunov Optimization

When employing Lyapunov optimization for compute offloading, it's critical to understand the drift and penalty. The penalty in an offloading scenario is usually the offloading aim, such as task dropping minimization or execution latency minimization, while the drift is usually the energy queue drift or task queue drift. By reducing the drift-plus-penalty expression, the best offloading decision and other parameters might be derived for each time frame. The Lyapunov optimization has a substantially lower computational complexity than (non-)convex optimization since the drift-plus-penalty equation is often a deterministic problem. Furthermore, unlike MDP, it does not require knowledge of the probability distribution of the random event process. The authors of [123] satisfied the requirements of latency and processing capability. To attain great long-term efficiency in online contexts, they created a UAV-Edge-Cloud computing model. They also employed Lyapunov optimization to keep track of the application's offloading costs and the environment's dynamics.

### 3.1.4 Markov Decision Process and Reinforcement Learning

The Markov Decision Process (MDP) is well suited to offloading optimization in situations where dynamic decision-making is needed due to changing environmental parameters such as wireless channel conditions and computation load. An MDP system normally consists of five-tuples ($\mathcal{K}, \mathcal{S}, \mathcal{A}, \wp, \mathcal{R}$), where $\mathcal{K}$ represents decision epochs; $\mathcal{S}$ denotes a set of states; $\mathcal{A}$ is the set of actions; $\wp$ represents the state transition probabilities, and $\mathcal{R}$ is reward. In this model, it is critical to define the five-tuple according to the offloading procedure [20]. Since Reinforcement Learning (RL) generally defines a problem as an MDP problem, the word "system definition" refers to how the system is defined in terms of state, actions, and reward. There is a central model in the RL process. Machine learning relies on input data and output effects, which are represented by the terms 'input' and 'output.''. Since RL can solve the curse of dimensionality problem posed by MDP, it is a perfect fit for a high-complexity offloading scenario like a large-scale UAV-MEC network.

Furthermore, since RL is an online tool, it is appropriate for use in a dynamic environment. In the work [124], the authors suggested multi-agent reinforcement learning (MARL) algorithms, in which two UAVs decide the goal helper and bandwidth allocation, to separate the assessment of the integrated decision. Similarly, to maximize the long-term utility of the proposed UAV-enabled MEC network, the aim of the optimization problem in [125] is to find the best computation offloading and resource management policies. The problem is further formulated as a semi-Markov method (SMDP), and deep RL based algorithms are proposed in both centralized and distributed UAV-enabled MEC networks, taking into account random system demands and time-varying communication channels. To support terrestrial infrastructures, the authors of [126] proposed a Non Orthogonal Multiplexing Access (NOMA)-enhanced UAV deployed in 3-D space. They formulate the sum rate maximization problem by optimizing the multi-dynamic UAV's trajectory and power allocation policy based on channel state information and user mobility. To tackle the formulated problem, they offered a two-step strategy. They employ the upper bounded K-means algorithm to determine user clusters on a regular basis. A multi-agent MDQN technique is developed based on the detected user association to jointly optimize UAVs' 3-D trajectory and power allocation policy to maximize overall throughput.

Having such a dynamic environment complicates the problem modelling in UAV-MEC. Offloading problems with several metrics are typically non-convex problems that are usually modelled using nonlinear integer programming (MINP) [127] [128] or MDP [129] or game theory [130] [131]. In reality, many recent works compound more than one modelling method and algorithms in one work to solve the complicated offloading decision problems as shown in table.4.

Table 4. Intelligent Algorithms for Computation offloading in UAV-MEC

| Ref | Optimization Technique(s) | Offloading Mode | Served terminals | Terminal Mobility | Computation efficiency | Energy efficiency | Time efficiency | Solution Algorithm(s) |
|---|---|---|---|---|---|---|---|---|
| [108] | convex optimization | partial offloading (sequential) | User Equipment (UE) | - | - | √ | √ | Upper Bound Search Algorithm |
| [132] | non-convex computation efficiency problem, | partial offloading (sequential) | Users | - | √ | √ | - | -Lagrangian dual method - the successive convex approximation (SCA) method |
| [133] | Non-cooperative Game +RL | partial offloading (sequential) | IoT | - | √ | - | - | -Best Response Dynamics - Binary Log-Linear (B-logit) and the Max Log-Linear (Max-logit) algorithms |
| [134] | a mix-integer non-convex optimization | Binary offloading | IoT | - | - | √ | - | block coordinate descent and successive convex approximation |
| [62] | non-convex problem | partial offloading (parallel) | Ground user | - | √ | √ | √ | block coordinate descent (BCD) method, successive convex approximation (SCA), and branch-and-cut method |
| [135] | Nonconvex optimization | Binary offloading | UE | - | - | √ | - | leveraging the Lagrange duality method, |
| [136] | Nonconvex optimization | partial offloading (parallel) | Users | - | - | √ | √ | Long Short Time Memory (LSTM)-based task prediction algorithm |
| [137] | MDP | Binary offloading | Users | - | - | - | √ | A2C |
| [119] | coalition formation game (CFG) and Stackelberg game | partial offloading (parallel) | Ground node | - | √ | - | - | -log-linear learning -The spatial adaptive play (SAP) |
| [138] | Non-cooperative Game | Binary offloading | mobile user (MU), | - | √ | - | - | UAV-enabled computation offloading (UECO) algorithm |
| [139] | Semi-MDP | Binary offloading | User | - | - | √ | - | Q-learning |
| [140] | non-convex problem | partial offloading (sequential) | IoT | | - | √ | - | iterative optimization algorithm with double-loop structure |

| Ref | | | | | | | |
|---|---|---|---|---|---|---|---|
| [141] | non-convex problem | partial offloading (parallel) | UE | - | √ | - | - | big-M reformulation, the penalized sequential convex programming, and the general Dinkelbach's method |
| [142] | RL | partial offloading (parallel) | UE | √ | - | - | √ | Deep Deterministic Policy Gradient (DDPG) |
| [143] | RL | partial offloading (parallel) | vehicle | √ | - | - | √ | DDPG |
| [144] | non-convex problem ( a two-layer joint optimization method) | | MDs | - | - | - | √ | Particle Swarm Optimization algorithm combined with Genetic Algorithm and greedy (PSO-GA-G) |

Table 4 shows that partial offloading is superior to binary offloading unless in IoT. The difficulties of non-convex offloading models are typically tackled using sequential convex approximation [132], the leveraging duality method [135], deep learning [136], or any tree search algorithm [133] [108], or the Particle Swarm Optimization algorithm and Genetic Algorithm [77]. On the other hand, RL algorithms such as Q-learning [139], DDPG [142], and A2C [137] are commonly used to solve MDP and RL models for UAV-MEC offloading problems. For the mobility of both MDs and UAV, the work in [142] [143] considers MDs and UAV mobility in a UAV-MEC system, but they approached the problem differently. The work [142] formulated the problem as minimizing the delay, while [143] formulated the problem as maximizing successful offloading tasks.

The intelligent algorithms used for solutions are choosen by the task's size and application time-sensitivity. In a dynamic environment like UAV-MEC, the degree of network structure (computation layers: MDs, UAV, and BS), such as the number of MDs, multi-UAV servers, and data offloading metrics for optimization object, are complicated issues. The offloading mode has also an impact on the complexity and behaviour of UAVs; table 5 shows how those factors are related.

Table 5. Offloading optimization objectives with modelling and solution

| Ref | Optimization objective(s) | Offloading modelling | Offloading mode | UAV structure and behaviour | Algorithms for solution |
|---|---|---|---|---|---|
| [145] | task's completion time | MINLP | Partial (parallel) | Multi-UAV (cooperative MEC) | Q-learning |
| [45] | Reducing the size of discarded tasks | MDP | Partial (Parallel) | One UAV | soft actor–critic (SAC) |
| [146] | maximize the total amount of processing tasks of all MDs | Non-convex (divide to 3 sub problem) | Partial (sequential) | Multi-UAV (cooperative MEC) | integer programming, convex optimization and successive convex approximation (SCA) |
| [147] | minimizing the total energy consumption of all MDs ( joint the offloading decisions of MD and the flight positions of UAV) | MDP | Partial (sequential) | Multi-UAV (cooperative MEC) | A3C |
| [148] | minimize energy consumption costs delay | non-convex quadratic constrained quadratic program | binary | Multi-UAV (cooperative MEC) | semi-definite relaxation |

| [149] | Minimizing computation latency and energy consumption | Stackelberg multi-layer game | Partial (Parallel) | Multi-UAV (cooperative MEC) | Stackelberg game algorithm (SGA) |

Table 5 shows that the solution of offloading problem is become more complex with multi-objective problems and multi-UAV network [149] [146] [147]. Also the offloading mode and formulated model is affects the complexity of solution algorithms so RL algorithms are the suitable solutions [145] [147].

## 3.2 Intelligent Price Decision for Computation Offloading in UAV-MEC

The offloading producers of the Space-Air-Ground Integrated Network depend on the MDs' positions in the network and the execution time of UAVs that can provide a variety of computing nodes choices. Some MDs in the network are in high-speed movement states. When processing high-intensity tasks, the signal strength of the wireless network coverage of the offloading UAVs will be weakened due to the long processing time [150].

The location of the offloading MDs and the UAV-MEC offloading node are not fixed. During the task offloading execution, the wireless channel performance from the MDs to the UAV servers in the space-based backbone network or the ground-based backbone network change. The longer the transmission delay is, the more difficult it is to guarantee the transmission quality of the wireless channel [151].

Therefore, it is necessary to fully consider the factors that affect the network status during the task offloading. To minimize the delay and energy consumption of task computation offloading, it is necessary to keep the current task offloading operation as the most optimal operation according to the announced price from UAV-MEC servers. When the MDs leave the coverage of the offloading server, which is currently executing the offloading task, or a new offloading server belonging to another operator that meets the offloading conditions, the MD changes the connection to the optimal UAV-MEC server, as illustrated in Fig.3. At this time, switching through the network should be considered to complete the task offload [67].

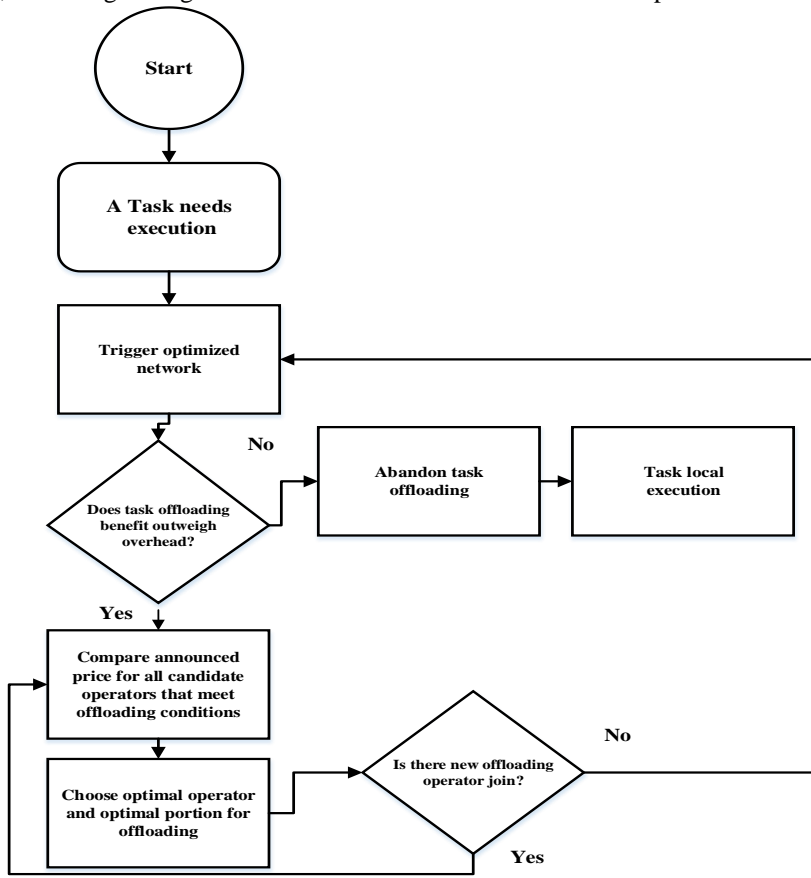

Figure 2. Price-offloading decision flowchart

The cost of offloading a task is determined by the cost of transmission to the UAV-MEC server, as well as the cost of executing the process [152]. As a result, various pricing methods exist depending on how the price target is set and who gets profits. Fig.3 depicts intelligent-based pricing decision flowchart. The MNO has a vested interest in ensuring that the flow quality meets the demands of its MDs. The wireless access connection, back-haul radio, and storage of content in UAV-MEC need the MNO to utilize UAV-MEC resources efficiently [14].

## 4 Intelligent Pricing Strategies in UAV-MEC

A future UAV system should be capable of self-planning trajectories and providing the finest computational capabilities to MDs. The interplay of UAVs and MDs, as well as UAV-UAV collaboration for long-term service provision and dynamic UAVs pricing, are all essential considerations [153]. The use of UAV-MECs to assist various ground agencies or terrestrial networks has recently been viewed as a critical component in the success of a large range of tasks that require significant improvements in terms of completion time, network performance, and flexibility. As a result, a well-organized definition of this paradigm must be precisely described while taking into account the various requirements for UAV-MECs [143]. The definition of the utility function in UAV-MEC has been adopted from the field of economics to reflect MDs' observed satisfaction from i) the enforced pricing policy by the MNO, (ii) resource allocation, and (iii) the fulfilment of their QoS prerequisites as Fig.4 illustrates. The purely intended fulfilment of the MD is expressed through the ratio of the data rate achievable to the corresponding uplink transmission power consumed. MD's satisfaction rises when the data rate and transmission power are both high. The latter leads to MD's battery life being extended and interference with other MDs being minimized [154].

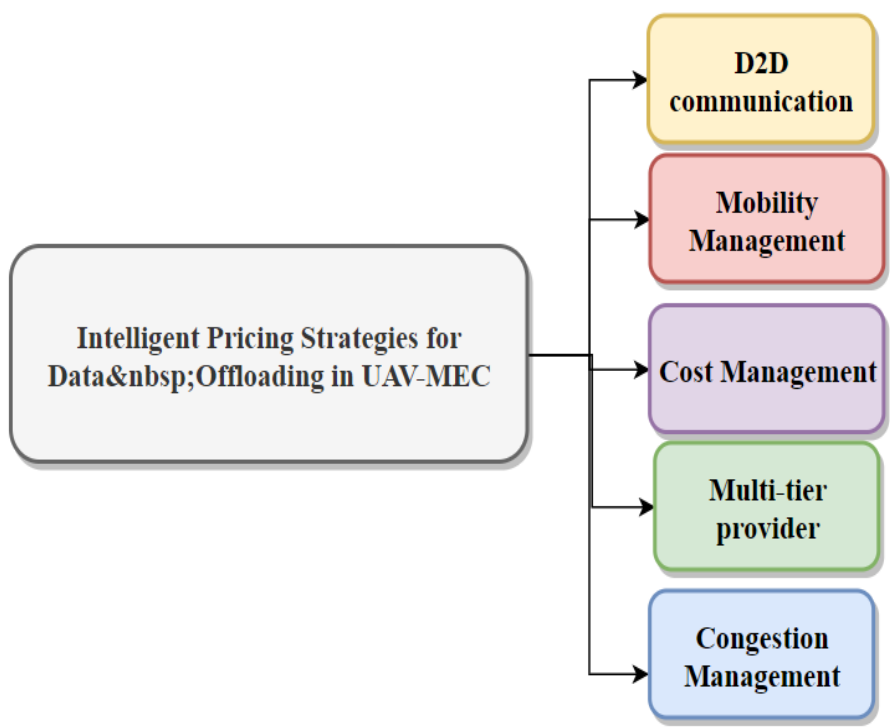

Figure 3.  Intelligent NUM Pricing Based strategies in UAV-ME

### 4.1 Price Strategy for D2D communication

Device-to-device (D2D) communication is effective in terms of transmission rate, energy cost, use of bandwidth, etc. in data offloading. However, unstable spectrum reuse performance can push MDs away, particularly when MDs are used as secondary or cognitive users of cellular networks [155]. The low expectations for D2D services and frequent mode switching can obstruct successful D2D offloading, cause severe network congestion, and result in poor QoE, particularly in densely populated areas [41]. Many authors focus on how to improve the use of D2D by encouraging the MD with idle resources to share its resource at a good price [156] [157]. To extend the V2V links to several cells and manage the inter-cell interference, the authors of [158] suggested an inter-cell V2V communication model with UAV assistance, in which a mutual UAV node is located in the middle of V2V. In [159] the cellular network makes money by charging V2V users an underlay spectrum access payment in exchange for co-channel interference from V2V connections. The interactions between the V2V links and the cellular links, which are the game follower and leader, are modelled using a Stackelberg game. The authors optimized the sum social community utility of social networking by power optimization, while all D2D and downlink users' QoS requirements must be met. Authors of [160] jointly integrated communication, computing, and caching resource utilization strategies to maximize UAV-MEC net revenue while ensuring MD QoE. To solve this joint optimization problem, a multi-dimensional hybrid adaptive particle swarm algorithm is proposed.

## 4.2   Price Strategy for Cost Management

For congestion relief and cost savings, the MNOs are trying to deploy UAV-MEC to offload MD data from their cellular networks. However, these network-centred approaches do not take into account the interests of MDs in monetary cost, energy usage, and deadlines for applications [161]. Offloading can only provide intermittent and opportunistic connectivity, resulting in non-negligible delay. With the rise of delay, users are increasingly becoming impatient, and their satisfaction would therefore be significantly reduced. Delayed traffic offloading is a promising model for allowing traffic with various delay characteristics and formulating the interrelationship between service delay and user satisfaction in a workable way [162]. It is a non-trivial and significant problem how the MD determines whether to offload their traffic to a complementary UAV-MEC. A competitive pricing strategy is therefore proposed to minimize the energy consumption of MDs under QoS constraints while maximizing the income of MNOs. To minimize the long-term average cost of computing in terms of power consumption and buffering delay at each MD [163] [164]. Moreover, for the User-Centric Problem in [165], a network centre's goal is to maximize the mutual value of all MDs, and MDs must be honest about their local information to find a solution. Authors showed that dual pricing, which is commonly used, could not ensure truthful information reporting from MDs unless the resource is oversupplied or the price is too high for the MD to afford.

## 4.3   Price Strategy for Mobility Management

Given MD mobility, UAV-MEC is considered a promising technology to support emerging applications in the vehicular network. Effective coordination between MDs and between MDs and UAV-MEC servers is needed to deal with the excessive computing power of the edge server and neighbouring MDs. The high mobility of MDs introduces difficulties in task offloading, scheduling, and reliable result feedback [166]. Thus, moving MD needs more resources to satisfy the required QoS, especially in the time-sensitive application, and that increases the cost paid by MNO to handle it. As a result, dynamic pricing is challenging to cope with mobility requirements and the trade-off between QoS and price of service [167] [168]. Hence, based on the MD mobility model and content caching model in [169], the MD association problem is conceived as a utility optimization problem, in which the QoE and handover expense of the MD are both taken into account. Furthermore, to solve the problem, the authors proposed an improved Dijkstra algorithm. In [170], the initial energy state of the UAV and the green energy charging feature is used to determine the moving UAV's relay forwarding and offload services. An MDP algorithm meets the needs of the realistic application scene and provides a versatile and efficient offloading mode with normalized device utility function maximization as the target.

## 4.4   Price Strategy for Congestion Management

Task offloading and resource allocation is performed separately by each MD to distribute the device load to minimize energy consumption. However, the offloading strategies of MDs on different layers can hardly be synchronized due to incomplete information, which can lead to network congestion [171]. Many works are proposed to reduce energy consumption [172] [173] [174] and delay management [175] [176] [177] by effectively regulating the offloading price. On the other hand, blockchain technology is a recent technology designed to improve network security and scalability, which has been applied in numerous aspects, including the Internet of Things (IoT) wide prospects. However, because IoT computation is confined in terms of communication and computing due to mobility and scalability, they are unable to endure the high cost of blockchain mining [178]. To overcome it, non-mining devices and edge clouds are used to construct a collaborative mining network (CMN) to conduct mobile blockchain mining jobs. Miners can outsource their mining tasks to non-mining devices inside a CMN or edge cloud (e.g., UAV-MEC) when there are inadequate resources and at a reasonable fee as illustrated in [179] [180] See table 6. Moreover, when a UAV-MEC encounters many hotspot candidates with varying user incidence rates and flight distances, the work in [181] demonstrated that using the UAV to support a single hotspot is the best option, taking into account the best pricing and energy allocation for each hotspot. However, when multiple UAVs are deployed on different hotspots, this result can be reversed, particularly when the hotspots are more symmetric or the number of UAVs is high. Since a UAV is battery-powered, it has a finite amount of energy for both mobility and communication. In addition, the work in [182] is allowing UAVs to move their radio modules to sleep mode to prolong battery life as an effective solution. This causes the contact function to be temporarily unavailable. The ultimate deal for a UAV operator in such a situation is to have a cost-effective service with reasonable availability.

## 4.5 Price Strategy for Multi-Tier Network

From the operator's point of view, in which a primary MNO rents its spectrum to a secondary Small Service Provider (SSP) like UAV-MEC network in return for some revenue from the SSP [183]. Since the UAV-MEC operator pushes the MNO with lower costs, MDs get a better service, as the UAV-MEC improves the spectrum efficiency of MDs and the resources available in the scenario. In addition, all MDs would prefer to offload data to the UAV-MEC server, and they will adapt their decisions. To optimize their respective market share, SSPs, like UAV operators [184] [68] need to agree on the best price strategy for task offloading. The offloading problem in [185] was investigated in a MEC network with two-tier UAVs. A MEC server is mounted on the High-Altitude Platform UAV (HAPUAV) to complete the computing tasks of the Low-Altitude Platform UAV (LAPUAV). To formulate the two-tier UAV MEC offloading problem, the authors proposed a multi-leader multi-follower Stackelberg game. To promote the incorporation of different UAVs into next-generation wireless communication networks, the authors of [186] suggested a hierarchical architecture of UAVs with multilayer and distributed functionality. In this regard, the work in [187] assessed the efficiency of smartphones and used the pervasive role of UAVs in disaster situations to help and speed up rescue operations. They suggested a cooperative communication scheme that takes advantage of existing network technology and accounts for different energy levels of multi-tiered network architecture to achieve a balanced energy consumption. Table.6 illustrates pricing strategies in UAV-MEC and solutions present.

Table 6 intelligent pricing strategies in UAV-MEC

| Ref | Pricing strategy | Market structure | | | how to formulate pricing | Solution |
| --- | --- | --- | --- | --- | --- | --- |
| | | seller | Buyer | Item | | |
| [188] | cost management | UAV-BS | Vehicle | AI model training | Maximum UAV coverage with lower cost | Gale-Shapley algorithm |
| [189] | congestion management | UAV-BS | Ground user | data rate | Maximize the served users with a certain data rate | pseudo-polynomial time |
| [190] | cost management | UAV-BS | Vehicle | Computational resources | Minimizing the energy consumption of UAV | Lyapunov optimization |
| [191] | cost management | MBS | UAV-BS | bandwidth | Maximize the offloading users | dynamic programming |
| [185] | congestion management | HAP-UAV | LAP-UAV | Computational resources | Minimizing delay of LAP-UAV | Stackelberg game |
| [192] | cost management | UAVs-cluster | mobile users | Computational resources | Minimizing average computation delay | long short term memory (LSTM) |
| [182] | cost management | UAV-BS | ground users | Computational resources | Minimizing energy of UAV-service provider | best response dynamics |
| [193] | cost management | UAV-BS | User | Computational resources | Maximizing the user offloading data | the best response dynamics |
| [157] | D2D communication | UAV-BS | Vehicle | Spectrum access | Minimizing the transmission power of V2V link | Stackelberg game |

| [181] | cost management | UAV-BS | User | Data rate and computational resources | Minimizing the energy of UAV | Dynamic programming |
|---|---|---|---|---|---|---|
| [194] | cost management | UAV-BS | user | Data rate | Minimizing power transmission of UAV | bisection search and concave-convex procedure (CCCP) methods |
| [180] | cost management | MNO | UAV operator | Communication resources | Minimizing UAV operator requested bandwidth | a Stackelberg game |
| [195] | cost management | Edge cloud | user | Computational resources | edge cloud utility maximization+ user utility maximization | Stackelberg game |
| [196] | cost management | Edge cloud | user | Computational resources | maximizing the benefits of edge clouds | Stackelberg game –Dynamic programming |
| [76] | cost management | UAV-BS | vehicle | Computational resources | maximizing the benefits of UAV | RL (DDPG) |
| [197] | cost management | Edge computing server | user | Computational resources | maximize the revenue of service provider | Duelling Double Deep Q Network (RL) |
| [198] | Mobility management | UAV-BS | Maritime IoT (M-IoT) | offloading transmission and computation | energy consumption minimization | DRL |
| [199] | Congestion management | MEC server | User | Computational resources | maximize the user's expected prospect-theoretic utility | semiautonomous game-theoretic , autonomous reinforcement learning-based approach |

According to Table 6, the UAV-MEC operator sells computational resources, spectrum access, data rate, and communication resources to MDs, whereas the UAV-MEC operator purchases communication and spectrum access resources from MNO. Since the MD requests good QoS, especially for delay-sensitive applications, as well as a cheap price, the pricing mechanism is challenging. From the standpoint of the operator, the pricing calculation is about lowering the cost of utilizing resources as well as the service latency to achieve MDs' QoS. The strategies used to tackle interactive problems are typically based on game theory [170] [157] [195], although the techniques for decision-making might range from Lyapunov optimization [190] to dynamic programming [182] [193] [181] [196] to deep learning [192] and reinforcement learning [76] [197] .

# 5  Open Issues

Despite the many benefits of using computation offloading and intelligent pricing strategies to maximize user and provider profit by offloading data to nearby nodes in UAV-MEC network, blindly applying intelligent computation offloading and intelligent pricing strategies is inadequate and more negotiation is needed. In this section, we categorize main barriers and high spot some research opportunities and future directions.

## 5.1  Network Scalability and Real Data Training

In the current literature, most of the proposed pricing learning algorithms for computation offloading are trained and evaluated on an offline simulator for their models. This makes it possible for a real-time implementation to use algorithm design and testing as a jump-starter. However, developing a simulator with sufficient fidelity to ensure that the learnt policy work can be carried out as envisaged when dealing with the real-world environment is incredibly challenging. The only way to get a definitive answer to the challenge appears to be to learn real-world price strategies [200]. Furthermore, researchers may use raw data to train the model from scratch, taking advantage of the current business model and adding advanced learning methods such as transfer learning and multi-objective learning. As a result, to increase AI performance, we consider offloading to edge intelligent servers such as UAV-MEC that have already been trained on similar tasks. The first challenge is to get several UAV-MEC servers to work together to provide AI services. The complexity of all types of offloading models is substantially increased by the large-scale network. As a result, the offloading decision time increases, increasing the overall offloading delay. Both Lyapunov optimization and machine learning are centralized methods that require data collecting. The collecting of data in a large-scale network may lead the network to become overloaded, compromising the real-time requirements of the offloading decision. This is because the choice is made only when all essential information has been received, and the time spent on this procedure is not insignificant. Furthermore, large-scale dynamic networks with strict real-time constraints, such as intelligent transportation systems or e-health infrastructures, are complex systems, in which the computation-offloading model must account for the heterogeneity of architectures and data. It's very critical to provide high-performance and scalable UAV swarm communication systems and routing protocols. Another challenge is how to deploy machine learning algorithms and how to train the network [201].

### 5.2 Competition or Cooperation of Multi-Operator Network

The ubiquitous wireless cellular network coverage/connectivity requirement has moved MNO's interest towards dense deployment of small cells with far narrower coverage areas compared to MBS. Today, small multi-operator cells belonging to different telecommunications operators are currently making effective use of resources, i.e., uninterrupted coverage and the ability to prevent under-use of network resources [202]. They work together to use caching techniques, store virtualization features, and edge computing to provide high-performance and large-scale computing services on demand [40]. However, the MNO will have little motivation to use its valuable resources to serve other MNOs' MDs unless it is reworded at a reasonable price. Many publications in the literature still encourage working in a competitive manner; however, others favour cooperative techniques to promote MD satisfaction. This issue still needs more investigation. UAV-MEC and a BS together also execute a large application in tandem. It should be noted that a large application can be broken down into several compute tasks. These tasks could be completed concurrently, but they must be completed in a specific order. Because this form of parallel computing can speed up the execution, figuring out how to reduce the time it takes to complete a collection of tasks is a critical challenge [203].

### 5.3 3D Mobility

MD mobility has a significant impact on the wireless communication channel between small cells and MD, so the recent UAV-MEC technology is favoured thanks to UAV channel gain. The question of whether or not this data should be offloaded to the new UAV-MEC server is a critical one. This problem arises as a result of a trade-off between the expense of offloading and the benefits of reduced delay and communication costs. Furthermore, for dynamic work offloading, migration, and optimal offloading decisions, a forecasting technology is necessary. All of these characteristics make offloading computations in mobility circumstances more complicated [20]. Because the UAV-MEC and MD are moving, the coverage is changed. To serve additional MDs and restrict the increase of its energy consumption, the UAV-MEC should locate an appropriate hover point and determine coverage at a fixed height [63]. The most important thing in a mobility environment is to strike a balance between delay and energy consumption. In delay-sensitive different applications, energy usage can be considered acceptable at a certain level so that the computational task can be offloaded and the outcomes can be installed with the shortest possible delay [21]. More research is still needed on price-based computation offloading for moving MDs [204]. The majority of works also made the assumption that the UAV is hovering at a constant height. Offloading while changing UAV height is still an unresolved research problem.

### 5.4 Security and Privacy

Blockchain is rapidly being used in several industries, including cryptocurrency, IoT [205], and AI [206]. The blockchain is designed to prevent malware nodes from stealing the price MDs paid to a higher level by ensuring safe, irreversible, and automated transactions between different entities. For more protection and privacy in UAV-MECs, there should be a degree of confidence and a high reputation score. It has been thought that resource-constrained UAV-MECs would not be able to support advanced protection algorithms (e.g., group signature). To fix this issue, the security functions of UAV-MECs should be offloaded to a UAV-MEC server, which performs these functions on their behalf. Due to the heterogeneous existence of UAV-MECs, which includes different types of communication protocols, complex security settings, and so on, this process requires robust security protection. Computation offloading and security functions of UAV-MEC servers pose a host of privacy issues. However, preserving the privacy of UAV-MECs is much more difficult than cloud computing [20]. However, with complicated cooperative interactions among UAV-MEC nodes and competition among MNOs, it is of great challenge to allocate both the security and wireless resource allocation in UAV-MEC networks. Moreover, UAVs could potentially be employed in military and defence applications due to their reliability and safety. UAVs can avoid obstacles, undertake surveillance, and pick and drop stuff, and their functionality is relatively straightforward to add and delete [207].

## 5.5 Artificial intelligent algorithms at the end MD (Federated Learning)

The advent of federated learning (FL) structures that will be adopted in future IoT systems is one of the most exciting distributed learning algorithms [208]. In FL, MDs can conduct a learning task cooperatively by simply offloading the local learning model to the UAV-MEC instead of sharing their entire training data [209]. Many researchers propose fully distributed learning running IoT nodes in the UAV-MEC as an agent for data training and decision-making reaction with the aid of the global entity node. The deep learning algorithm requires an appropriate computational resource that could not be trained and tested by the IoT sensor in those days. Thus, what is the cost of fully distributed learning and what portion of computing resources need to be offloaded to UAV-MEC. Because there are so many AI approaches that can be used for different applications, it's difficult for even a professional researcher to pick the right one. Because the UAV-MEC communication system is a multi-dimensional network that is more complicated than current terrestrial communication networks, the research community must continue to investigate how to develop an effective AI technique. When we implement AI-based solutions, on the other hand, the calculation time and transmission latency between the UAV-MEC and the ground station will have a detrimental impact on network performance. Because AI approaches typically require significantly more computations than traditional methods, it is vital to increase computation efficiency when using AI to improve the performance of a UAV-MEC communication network [210].

## 5.6 UAV Coalition

Future inter- and intra-UAV-MEC communication technology will be hard up by technical improvement. LoRa and 6LoWPAN have also arisen as favourable technologies for UAV-MEC communication over small distances. Disruption of frequency, rate adaptation, high altitude performance, and versatility are all difficulties [211]. The majority of previous computation offloading research has concluded that the computation tasks are homogeneous. The computation offloading method is made simpler by this assumption. In reality, there are a variety of activities. Some tasks, for example, are pre-emptive (i.e., they will pre-empt other tasks), whereas others are not. The difficulty of offloading is greatly increased by the heterogeneity of tasks. The computing units may also be the source of heterogeneity. As a consequence, a single UAV-MEC can be unable to provide computation offloading for multiple tasks [20]. MDs and UAV-MEC servers, for starter, form numerous coalitions. MDs are allocated to a UAV-MEC server within the same coalition for computation offloading. The proposed definition of "coverage chance," which suggests the availability of UAV-MEC servers to MDs, is inextricably linked to UAV-MEC server selection. Furthermore, a pricing system is used to encourage MDs and MEC servers to work together [212]. Specifically, collision avoidance is an important feature of the UAV-MEC coalition, as is learning how the UAV-MECs can adjust their routes to avoid collisions while tracking and hitting the target. It will be fascinating to see how the multiple UAV-MEC deploys for the mission perform in such a scenario [213].

## 5.7 5G Communication Methods for UAV-MEC

In terms of energy savings, rapid integration, and adoption ease, the use of 5G and modern networking methods such as Non-Orthogonal Multiple Access (NOMA), Industrial IoUAV, and others has shown favorable results. Academic and industrial researchers, on the other hand, are currently investigating accurate models for cellular-connected UAV-MEC networks using a variety of performances [214]. However, there is still a scarcity of research on computation offloading with SDN provision. To offload to SDN, the computational and control operations must be segregated. Although computation offloading in SDN linked UAV-MEC is still in its early phases, further study is needed to address several outstanding challenges, particularly in offloading for both centralized and distributed SDNs, taking into consideration new characteristics such as data timeliness and authentication [215]. Furthermore, several situations in the mmWave 5G setting would almost certainly include buildings as well as wind perturbations. This would necessitate simulations that account for the blockage effects and perturbations in the UAV-MEC's flight path. The platform's throughput, latency, and subsequent consequences for stability can then be investigated further [216]. Another issue is delay-aware scheduling for multiplexing techniques; however, the work [217] only considers delay-aware scheduling in an orthogonal frequency division multiple access (OFDMA) MEC system; other techniques require more research. In addition, in a 5G network, UAV-MEC will use a Cognitive Radio Network (CRN) to interact with the main MNO by accessing spectrum gaps without interfering with the active primary user of the MNO [218]. Another issue is beamforming, it is important to think about how the flying direction of UAV-MEC affects the beamforming design. This is difficult since the UAV's direction varies dynamically over time due to UAV mobility. The design of dynamic beamforming for UAV-MEC networks necessitates substantial research [219].

### 5.8 UAV Trajectory Optimization and Flying-Time

UAV-MECs are a promising technology due to flexible wireless connectivity and coverage even in the absence of network infrastructure. Despite the fact that UAV-MECs have numerous applications ranging from mobile relay BS to caching and MEC cloudlets, it is critical to thoroughly investigate UAV trajectory optimization [220] [221] [222], hovering altitude [223] [224], and speed control [225] [226]. The UAV-MEC mobility causes them to leave the coverage area of the serviced MDs, which may increase latency [227]; additionally, there is a large amount of data to be offloaded from MDs concerning the available bandwidth on both UAV-MECS and the backhaul network [228]. This movement had a significant impact on QoS, so the mobility of UAV-MEC needs to plan carefully. Another issue is UAV-enabled WPT [229]; unlike traditional ground-station Electrical Transmitters (ETs), UAVs can act as a new type of aerial ET that can fly freely to efficiently charge adjacent low-power devices and other UAVs [230]. The charging process occurs on a regular basis; however, each UAV-MEC consumes different amounts of energy. This difference in energy consumption necessitates more investigation.

### 5.9 Multi-objective problems in UAV-MEC Networks

The UAV typically functions as BS with a MEC server, MDs terminals, or a relay in the complicated UAV-MEC 5G network. The network includes devices like ground-based BS, a Marco-Cell controller, and UAVs that are MEC or relay or IoT devices [231]. Due to the complexity of the environment and the quantity of UAVs, BS, and MDs, the performance measures employed as a goal for offloading are typically constrained. The UAV-MEC network becomes increasingly complex due to the mobility of both MDs and UAVs. The complexity of the network increases as there are more optimization objectives. Even though there are many MDs with more than two offloading optimization metrics and few works that are studied offloading with mobility of both multi-UAV and MDs. The Complexity of dynamic pricing for the computation offloading process with multi-objective functions in UAV-MEC is challenging and this issue still needs further study.

## 6 Conclusion

In UAV-MEC networks, computation offloading plays a key role in cooperative and collaborative network operations. Multiple UAV-MECs can carry out complex tasks efficiently when they are organized as an ad hoc network or assisted cellular network, where wireless communication in a cellular network is essential for cooperation and collaboration between UAV-MECs and the ground station. UAV-MEC provides a promising approach to significantly reduce network operating costs and improve QoS of MDs by aggressive computation resources to the network edges. UAVs will become more widely employed for AI services. This paper surveyed and analysed the works related to computation offloading advantages in UAV-MEC. In this regard, a detailed analysis of intelligent computation offloading process and factors has been identified. Since the intelligent computation offloading, and dynamic pricing in this area is critical, the dynamic pricing strategies in UAV-MEC networks were analysed. Finally, the survey painted some research directions and challenges, among which computation offloading in UAV-MEC should be the key area that needs further investigation. This survey outlined open issues as well as future research directions.

The UAV-MECs will be used not just for 5G and beyond, but also as a satellite communications relay in the future. UAV-MECs are not only for offloading data from MDs but also for charging MDs, notably IoT and charging UAV-MECs that are utilized for offloading cellular networks' data. UAV-MEC is a promising technology with a wide range of applications, including military, rescue, and object delivery.